\documentclass[aps,pre,twocolumn]{revtex4-1}

\usepackage{graphicx}

\usepackage{enumerate}

\usepackage{amssymb,amsfonts,amsmath,color}

\newcommand{\SetS}{\boldsymbol{\mathcal{S}}}

\begin{document}

\title{Alternative steady states in random ecological networks}

\author{Yael Fried}
\affiliation{Department of Physics, Bar-Ilan University,
Ramat-Gan IL52900, Israel}

\author{Nadav M. Shnerb}
\affiliation{Department of Physics, Bar-Ilan University,
Ramat-Gan IL52900, Israel}

\author{David A. Kessler}
\affiliation{Department of Physics, Bar-Ilan University,
Ramat-Gan IL52900, Israel}


\begin{abstract}
\noindent In many natural situations one observes a local system with many competing species which is coupled by weak immigration to  a regional species pool. The dynamics of such a system is dominated by its stable and uninvadable (SU) states.  When the competition matrix is random, the number of SUs depends on the average value of its entries and the variance. Here we consider the problem in the limit of weak competition and large variance. Using a yes/no interaction model, we show that the number of SUs corresponds to the number of maximum cliques in a network close to its fully connected limit. The number of SUs grows exponentially with the number of species in this limit, unless the network is completely asymmetric. In the asymmetric limit the number of SUs is ${\cal O} (1)$. Numerical simulations suggest that these results are valid for models with continuous distribution of competition terms.
\end{abstract}

\maketitle

\section{Introduction}

The richness of ecological communities poses a prolonged theoretical challenge. Focusing on guilds of many species competing for a common resource (and neglecting, for the moment, processes like predation or mutualism) the main problems are two. First, the competitive exclusion principle \cite{gause1934struggle, hardin1960competitive} suggests that the result of competition for a single limiting resource is the extinction of all  except the fittest speices, and that in the presence of a few limiting resources the equilibrium number of species is smaller than or equal  to  the number of resources \cite{tilman1982resource}. Second, even if the number of limiting resources is large, May \cite{may1972will} pointed out that if the niche overlap between species is substantial the chance of a system of $N$ species to admit a  stable equilibrium decreases exponentially with $N$.  May's result is based on the spectral properties of random stability matrices. Practically, it  implies that to achieve a stable coexistence  of more than 6-8 species one has to fine-tune the competition parameters in an unrealistic manner.

However, in many  situations the (inter and intra) specific  dynamics takes place on local patches, which are coupled by migration to each other or to a regional  pool. Accordingly, many ecological models are focused on the dynamics of a local patch, putting aside the global stability problem. A \emph{mainland-island} model \cite{macarthur1967theory,bierregaard2010losos} is the  simplest scenario considered in this context: a set of local populations of different species are competing with each other and the island is exposed to weak migration from a static  pool  of $N$ species on the mainland. The structure of the community on the island reflects a balance between local extinctions  and colonization by immigrants from the mainland. Extinctions may be either deterministic, due to the pressure that a species suffers from its competitors (or from the local environment), or stochastic, caused by the random nature of the birth-death process \cite{kessler2007extinction,ovaskainen2010stochastic} possibly superimposed on the effect of environmental variations \cite{lande2003stochastic,danino2016effect}.

In a recent work \cite{kessler2015generalized}, Kessler and Shnerb suggested a classification of the qualitative features of the community on the island. Four different ``phases" were identified.

\begin{enumerate}[I.]
  \item \textbf{Full coexistence}: If the interspecific competition is weak (say, if different species use essentially different resources) any species in the mainland may invade the island and establish a finite population, so all $N$ species are present on the island. Local extinctions still occur but if the local populations in steady state are large, these events are rare and transient. Technically speaking, the deterministic (i.e., noise-free) model allows for a stable fixed point with all the $N$ species coexisting.
  \item \textbf{Partial coexistence}: As the competition among species grows, the species' abundances decay as they feel more pressure from other species. Since the competition matrix is heterogenous, some species feel more pressure than others, and  the deterministic model eventually allows for a stable fixed point for a finite subset containing $S$ of the $N$ species. The other $N-S$ species on the mainland cannot invade the island, i.e., their growth rate at low densities on the island is negative.

      \item \textbf{Disordered}: When the competition increases even further, \emph{and the competition matrix is not symmetric }(meaning that species 1 may put a lot of pressure on species 2 but species 2 puts much less pressure on species 1, say), the system may not have an attractive fixed point at all or, even if it have one, its basin of attraction will be very narrow. In the presence of noise, the system fails to converge to an equilibrium state and instead it shows intermittent behavior with many long excursions that reflect high-dimensional chaotic/periodic trajectories.

      \item \textbf{Alternative steady states}: Finally, when the competition terms are  large, there will be a number of different subsets of  the $N$ species which are both stable and uninvadable. For example, if the interspecific competition is extremely large and the island is colonized by a single species, all other mainland species cannot invade, so one needs to wait for a rare stochastic extinction in order to see species turnover. If the competition is not so strong, subsets of more than one species play the same role: species within the subset interact only weakly so they may live happily together, but other species cannot invade.
\end{enumerate}

The aim of this paper is to discuss this last phase, which is characterized by strong competition and alternative steady states. The immediate motivation for this discussion comes from a recent paper by Fisher and Mehta \cite{fisher2014transition}, who suggested that in this phase the dynamics of the island exhibits a glass transition: with weak noise/immigration the system is trapped  for most of the time in one of the  SUs, while when  stochastic effects are strong it behaves like a ``liquid" and its dynamics is closer to the disordered phase discussed above. In \cite{fisher2014transition}, a version of the symmetric competition model with strong interaction was mapped into a well-known physical model for glassy behavior, the random energy model \cite{derrida1981random}.

Technically, the appearance of a  glass transition in the random energy model is related to the exponential increase of the number of local minima with the system size (here, the number of mainland species, $N$). When both the energy and the entropy increase linearly with the system size, a glass transition appears at finite temperature (level of noise). Therefore, it is natural to investigate the scaling of the  number of SUs on an island with the number of species. In fact, this problem is considered by ecologists for many years \cite{gilpin1976multiple,Lischke2017}.

Recently, we have studied this problem and derived a few exact results \cite{fried2016communities}. Using a version of the model we call the Binomial model, we have mapped the problem of counting SUs to that of finding  the number of maximal cliques in a random network. We showed that in a particular parameter regime the number of SUs is \emph{not} exponential; it goes like $N^{\ln(N)}$ for symmetric networks, and like $N/\ln^{3/2}(N)$ for (fully) asymmetric networks.

Here we are going to analyze the very same model in a different parameter regime, which includes the case  where the competition is weak and the heterogeneity is strong (this is the case considered in \cite{fisher2014transition} and \cite{bunin2016interaction}, see discussion). We will show that in this regime the number of SUs indeed increases exponentially with $N$,  if the system is not fully asymmetric. On the other hand, for the asymmetric system (see the definitions below) the number of SUs in this regime is order one. We also show how to make a connection between this weak competition regime and the results obtained for  strong competition in \cite{fried2016communities}, and provide some intuitive argument.

This paper is organized as follows. In the next section we summarized the results of \cite{fried2016communities}; we present the  generalized competitive Lotka-Volterra model (GCLV) and our simplified,  Binomial model, and show how to map SUs to  maximal cliques, arriving at the formula of Bollob{\'a}s and Erd{\"o}s~\cite{bollobas1976cliques}. In the next two sections we present our main result,  the number of SUs, as calculated from this formula, for the symmetric and the asymmetric case. We also present numerical computations  showing that the results of the Binomial model also describe the qualitative behavior of the more realistic Gamma model. Finally we discuss the works of Refs.~\cite{fisher2014transition} and \cite{bunin2016interaction} and the relevance of the results presented herein and in these papers to realistic ecological networks.

\section{The model}

To get oriented, we start with a system of two competing species without noise and immigration. The GCLV reads
 \begin{eqnarray} \label{eqa1a}
\frac{dx_1}{dt} = x_1-x_1(x_1+ {\tilde c}_{1,2}  x_2) \\ \nonumber
\frac{dx_2}{dt} = x_2-x_2(x_2+ {\tilde c}_{2,1}  x_1),
\end{eqnarray}
where $x_i$ is the abundance of each of the species. This system is characterized by the competition matrix
 \[  \left( \begin{array}{cc}
0 & {\tilde c}_{1,2}   \\
{\tilde c}_{2,1} & 0  \\
 \end{array} \right).\]
 where the intraspecific density dependence (a decrease in the growth rate with abundance, manifested in the diagonal term) was taken to be one and is not part of the matrix. The stress put upon species 1 by species 2 is ${\tilde c}_{1,2}$ and the stress put upon 2 by 1 is $ {\tilde c}_{2,1}$. $\rho \equiv {\tilde c}_{1,2}+{\tilde c}_{2,1}$ is a rough measure for the niche overlap, or total strength of competition in the system. $\kappa \equiv {\tilde c}_{1,2}-{\tilde c}_{2,1}$ measures the heterogeneity of the competition matrix, i.e., it tells us how much the species differ from each other in their response to an increase in the density of a competitor. We consider a model as \emph{symmetric} if ${\tilde c}_{i,j}={\tilde c}_{j,i}$ for any pair of species, and as \emph{asymmetric} if there is no correlation between ${\tilde c}_{i,j}$ and ${\tilde c}_{j,i}$.

 A steady solution for  (\ref{eqa1a}) in which both $x_1$ and $x_2$ are non-negative is called a feasible solution (we cannot allow negative densities). If both densities are positive and the solution is stable, we called it a coexistence solution. Such a solution for (\ref{eqa1a}) exists as long as $\rho <2-|\kappa|$, meaning that, for a given level of niche overlap $\rho$ the system becomes less stable as the heterogeneity grows. This basic logic holds also in more diverse systems.

For a system of many competing species the GCLV is:
\begin{equation} \label{eqa2}
\frac{dx_i}{dt} = x_i-x_i^2-\sum\limits_{j \neq i} {\tilde c}_{i,j} x_i x_j.
\end{equation}
The mean of the  terms of the competition matrix,
$$ C \equiv \frac{1}{N(N-1)} \sum\limits_{i,j} {\tilde c}_{i,j},$$ reflects the overall strength of the competition
in the system. The variance of these entries, $\tilde{\sigma}^2$, is the simplest measure for its heterogeneity.  To emphasize these properties we will factor out the average from the competition matrix, so the GCLV takes the form,
\begin{equation} \label{eqa3a}
\frac{dx_i}{dt} = x_i - x_i \left(x_i + C \sum\limits_{j \neq i }^N c_{i,j} x_j \right),
\end{equation}
 where $\overline{c}_{i,j}=1$.

May's analysis \cite{may1972will} of the complexity-stability problem is based on the observation that a linear stability features of a feasible solution of (\ref{eqa3a}) yields an $N \times N$ random matrix which is similar to the interaction matrix. For such a state to be stable all the eigenvalues of this matrix should be negative. However, a random matrix with $(-1)$ on the diagonal and off diagonal terms with mean zero and variance $C^2 \tilde{\sigma}^2$
has its eigenvalues between $-1+C\tilde{\sigma} \sqrt{N}$ and $-1 - C\tilde{\sigma} \sqrt{N}$, so  a feasible  solution for  (\ref{eqa3a}) is almost surely unstable when $N$, the number of species, is above $N_c =1/(C  \sigma)^2$. The applicability of this argument to purely competitive systems requires some more discussion, since the main problem in these systems is to ensure feasibility \cite{rozdilsky2001complexity,kessler2015generalized}, but the main insight turns out to be valid here as well.

In this paper, as in \cite{fried2016communities}, we are interested in the features of the system way above this ``May limit", i.e.,  when $N \gg N_c$ and the system supports alternative steady states. What we would like to know is how many stable and uninvadable (SU) subsets of the $N$ species exist, i.e.,  how many  $S$-subsets of the $N$ species satisfy the following two conditions:

\begin{enumerate}
  \item \emph{Stability and feasibility:}  Eq. (\ref{eqa3a}), \emph{when limited to a specific size $S$ subset, $\SetS$}, yields a time independent solution for which ${\bar x}_i >0$ for all of the species in $\SetS$, where ${\bar x}_i$ is the equilibrium density of the $i$-th species in the subcommunity.

  \item \emph{Uninvadability:}  Eq. (\ref{eqa3a}), when applied to all absent $N-S$ species and linearized around the fixed point $x_i = {\bar x}_i$ for $i \in \SetS$ and $x_i = 0$ for $i \not\in \SetS$, yields negative growth rates ${\dot x}_i/x_i$ for all $i \not\in \SetS$.
\end{enumerate}

 We are interested in the SU enumeration problem for a random matrix, so we would like to draw the $c_{i,j}$s from a uniform, positive semi-definite, distribution with a mean one and a given variance~\cite{fisher2014transition,kessler2015generalized}. For our numerics we have used the Gamma distribution for this purpose, and denote this as the Gamma model.   A $c_{i,j}$ matrix (for simplicity the examples are given for the symmetric case) may look like,
 \[  \left( \begin{array}{cccc}
0 & 0.95 & 1.63 & 0.96 \\
0.95 & 0 & 0.48 & 0.97 \\
1.63 & 0.48 & 0 & 1.12 \\
0.96 & 0.97 & 1.12 &0  \end{array} \right).\].

 To map this model to the maximum clique problem,  we treat an alternative model, the Binomial (yes/no) model, where all the elements of the $c_{i,j}$ matrix (in the asymmetric case; the pair $c_{i,j}=c_{j,i}$ in the symmetric case) either are strictly zero (with probability $p$) or (with probability $1-p$) are equal to a finite constant $C \cdot A$, so the interaction matrix $\tilde{c}_{i,j}$ takes the form, say,
  \[ C \left( \begin{array}{cccc}
0 & A & 0 & A \\
A & 0 & 0 & A \\
0 & 0 & 0 & 0 \\
A & A & 0 &0   \end{array} \right).\]

The Gamma and the Binomial model have the same competition strength, $C$, if
\begin{equation}
A = \frac{1}{1-p}.
\end{equation}
The variance  of the matrix elements of the Binomial model is given by,
\begin{equation}
\tilde{\sigma}^2 =  C^2 \frac{p}{1-p}.
\end{equation}

For the symmetric model, if $C$ is large enough, each pair of species $i$ and $j$ is either non-interfering, ${\tilde c}_{i,j}={\tilde c}_{j,i}=0$, or mutually exclusive.    Accordingly, as explained in \cite{fried2016communities}, the SU problem has a geometrical interpretation. For a graph in which each node represent a species and each pair of non-interfering species is connected by an edge, a stable state corresponds to a subset $\SetS$ of nodes such that the induced subgraph is complete. For this stable state to be uninvadable  any vertex that is not a part of the clique is required to  have at least one mutually exclusive species in the clique, i.e., that the clique is maximal such that it cannot be extended by including any other connected vertex. Accordingly, for large $C$ the number of SUs is equal to the number of maximal cliques of the corresponding graph.

In \cite{fried2016communities} we showed that, as long as $p$ is ${\cal O} (1)$, the growth of the  number of maximal cliques, $SU(N)$ with $N$
 is \emph{not} exponential, and in fact for a symmetric system it grows as
 \begin{equation} \label{eqa0}
SU(N) \sim N^{\zeta(p) \ln(N)},
 \end{equation}
 where $\zeta(p) = 1/[2\ln(1/p)].$  Clearly, the expression (\ref{eqa0}) must fail when the value of $p$ is close to one, since  $p \to 1$ implies $\ln(1/p) \to 0$ and $ SU(N) \to \infty$. On the other hand when $p=1$ we reach an extreme stabilization and the system has only one stable uninvadable state, $SU(N) = 1$.  In order to clarify the behavior of the system in this limit, in the next two sections we will find a formula for  $SU(N)$ in the limit $N\to \infty$, where $p=1-\alpha/N$ and $\alpha={\cal{O}}(1)$.

\section{The symmetric case}

In this section we consider the symmetric version of the binomial model. Every pair of species is noninterfering ($c_{i,j}=c_{j,i}=0$) with probability $p$ and have symmetric competition  ($c_{i,j}=c_{j,i}= A$) with probability $1-p$. We assume that $C$ is large such that no pair of competing species is allowed on the island (if they are interacting, then they are mutually exclusive), and a species cannot invade the island in the presence of  one of its competitors. Accordingly, each maximal clique of the network is a stable and uninvadable state.

To get the basic intuition for the results we derive in this section, let us consider the case $p=1$, i.e., all species are noninteracting. Clearly, in this case there is only one maximal clique - the one with all the $N$ species.

Now let us break the link between, say, species 1 and 2, so $c_{1,2}=c_{2,1}= A$. The number of maximal cliques is now two: species 1 and all the species $3..N$ and the set $2..N$. Breaking the next link (without loss of generality, between 3 and 4) doubles the number of maximal cliques and so on. Hence, the number of maximal cliques grows exponentially as $p$ decreases, until we start to break more than one link per node. Since there are ${\cal O} (N^2)$ links in the system, this will happen when the number of broken links is ${\cal O} (N)$. Accordingly, one expects that, when the deviation of $p$ from one is ${\cal O} (1/N)$, the number of SUs will be exponentially large in $N$.

Bollob{\'a}s \& Erd{\"o}s~\cite{bollobas1976cliques} showed that the number of maximal cliques of size $S$ in a random graph of $N$ nodes is given by
\begin{equation} \label{eqa1}
SU(N,S) = {N \choose S} p^{S(S-1)/2} (1-p^S)^{N-S}.
\end{equation}
In \cite{fried2016communities}, we performed the sum over $S$, giving the behavior of $SU(N)$, Eq. (\ref{eqa0}), when $p$ is not too close to unity.

 Now let us find the leading asymptotic behavior of the  sum (\ref{eqa1}) when $1-p$ is small, of order ${\cal{O}}(1/N)$. More precisely, we  define
\begin{equation}
p \equiv 1- \frac{\alpha}{N},  \quad       S\equiv N\beta,
\end{equation}
where $\alpha$ and  $\beta$ are both  ${\cal{O}}(1)$. With these definitions, (\ref{eqa1})
reads
\begin{widetext}
\begin{equation} \label{eqa2}
SU(N,\beta) = {N \choose N\beta} \left(1-\frac{\alpha}{N} \right)^{N\beta(N\beta-1)/2} \left(1-(1-\frac{\alpha}{N})^{N\beta} \right)^{N-N\beta}.
\end{equation}
\end{widetext}
In the large $N$  limit,  the expression  (\ref{eqa2}) may be written as,
\begin{equation} \label{eqa3}
SU(N, \beta) \approx {N \choose N\beta} e^{-N\alpha\beta^2/2} e^{N(1-\beta)\ln(1-e^{-\alpha\beta})}
\end{equation}
Since for $\beta\sim{\cal{O}}(1)$, both $N\beta$ and $N(1-\beta)$are large, we can approximate the combinatorial factor using Stirling's formula, giving

\begin{equation}
{N \choose N\beta} \approx
\frac{e^{-N\left[(1-\beta)\ln(1-\beta)-\beta \ln(\beta)\right]}}{\sqrt{2\pi N \beta (1-\beta)}}.\end{equation}
Accordingly,
\begin{equation} \label{eqa5}
SU(N,\beta)\approx \frac{e^{Nf(\beta)}}{\sqrt{2\pi N\beta(1-\beta}},
\end{equation}
where

\begin{eqnarray}\label{eqa7}
F(\beta)&=&-(1-\beta)\ln(1-\beta)-\beta \ln(\beta) \\ \nonumber &-&\frac{\alpha\beta^2}{2}+(1-\beta)\ln(1-e^{-\alpha\beta}).
\end{eqnarray}

The total number of SUs is the sum over $S$ of $SU(N,S)$, which is translated to an integral over $\beta$ of (\ref{eqa5}). This integral may be approximated via Laplace's method, as $F(\beta)$ has a maximum in the range $0<\beta<1$. We denote the location of this maximum by $\beta_0$, which depends on $\alpha$ and satisfies

\begin{eqnarray} \label{eq23}
F'(\beta_0) &=&  -\alpha\beta_0+ \frac{\alpha(1-\beta_0)}{e^{\alpha\beta_0}-1} \\ \nonumber &+&
\ln(1-\beta_0)-\ln(\beta_0)-\ln(1-e^{-\alpha\beta_0}) = 0.
\end{eqnarray}

\begin{figure}
\includegraphics[width=7cm]{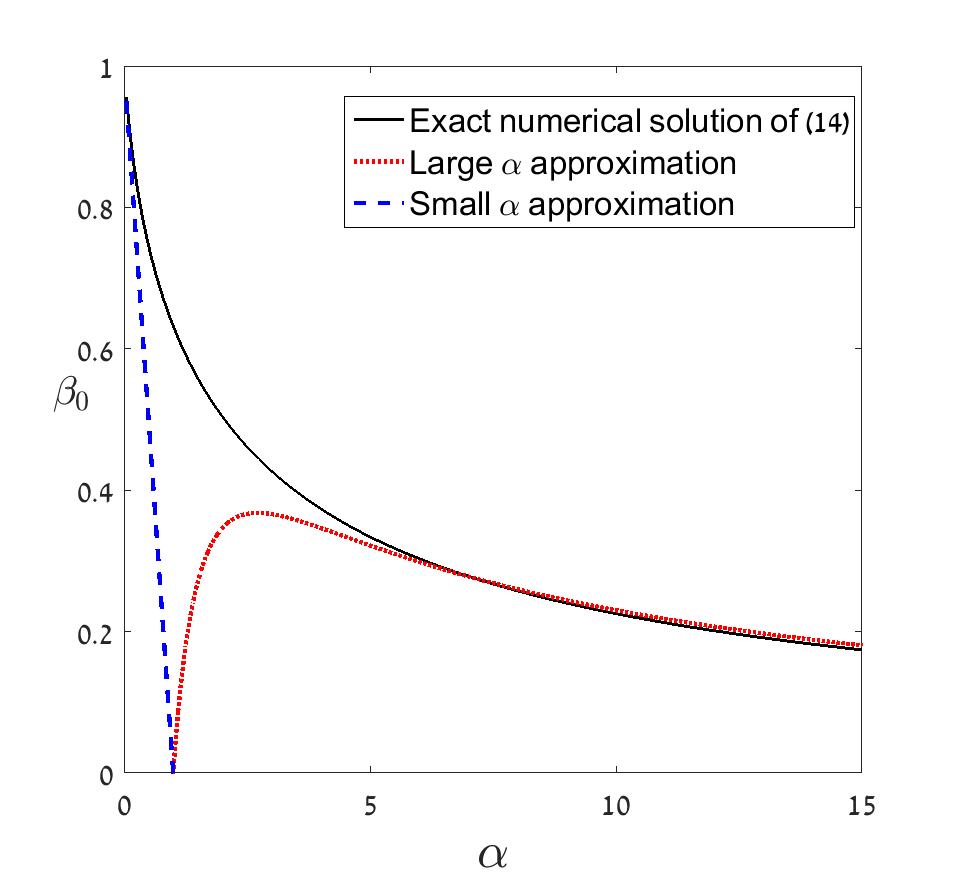}
\caption{Solutions for Eq. (\ref{eq23}). The full, black line is the exact numerical solution, the dashed blue is $1-\alpha$ and the dotted red line depicts $ \ln(\alpha)/\alpha$.   }\label{fig1}
\end{figure}

\begin{figure}
\includegraphics[width=7cm]{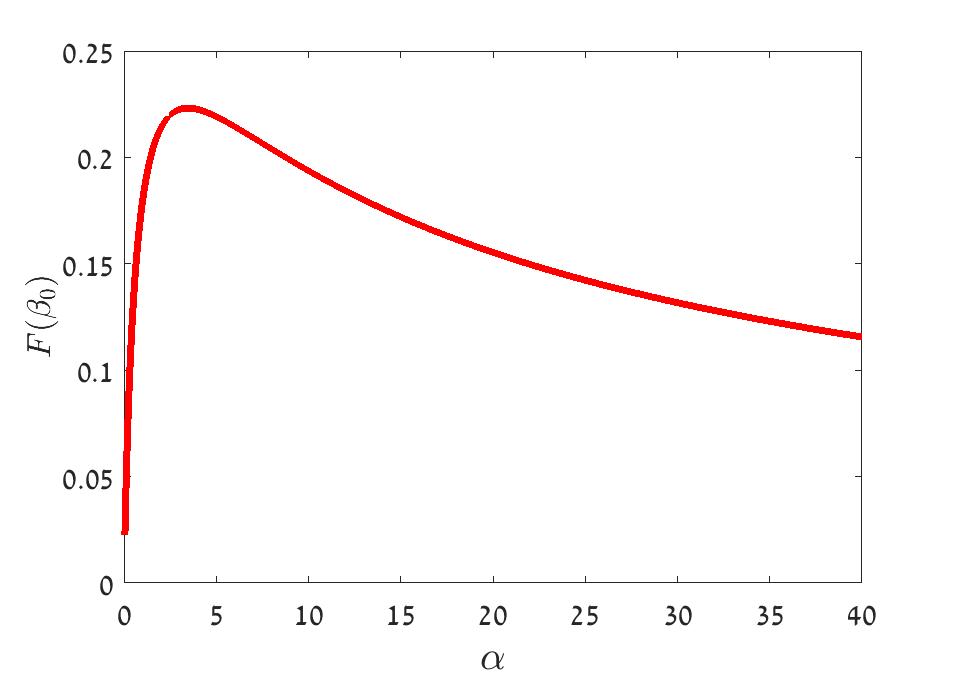}
\caption{$F(\beta_0(\alpha))$, or $\ln(SU)/N$, vs. $\alpha$ for $0<\alpha<40$. The number of SUs grows exponentially with $N$, where the coefficient of the exponent is between $0$ and $0.25$.    }\label{fig2}
\end{figure}

The graph of $\beta_0(\alpha)$ is depicted in Fig. \ref{fig1}. Then, to leading order, the total number of SUs, $SU(N)$, is given to leading order by
\begin{equation} \label{eqa8}
SU(N) \equiv N \int_0^1 SU(N,\beta) d \beta  = \frac{ e^{NF(\beta_0)}}{N \sqrt{|F''(\beta_0)|}\beta_0(1-\beta_0)},
\end{equation}
so that indeed the number of SUs increasing exponentially with $N$ in this parameter range.
Here we will have an interest only in the controlling factor, so we focus on $F(\beta_0)$. For general $\alpha$, this needs to computed numerically, with the results shown in Fig. \ref{fig2}. We see that $F(\beta_0)$ rises from 0 at $\alpha=0$, reaches a maximum and then decreases slowly for large $\alpha$.  The behavior at large and small $\alpha$ is accessible to analysis.  For small $\alpha$, we see from Fig. \ref{fig1} that $\beta_0$ is close to unity and thus to leading order in $\alpha$, $1-\beta$ we have
\begin{equation}
F'(\beta_0) \approx \ln(1-\beta_0) - \ln \alpha
\end{equation}
so that $\beta\approx 1-\alpha$ and $F(\beta_0) \approx \alpha/2$.  For large $\alpha$, since $\beta_0$ is small but $\alpha\beta_0$ is large,  the dominant balance of terms for large $\alpha$ is
\begin{equation}
 -\alpha\beta_0+ \frac{\alpha}{e^{\alpha\beta_0}-1}\approx
\ln(\beta_0).
\end{equation}
We can exactly solve this equation using an auxiliary variable $r$, $\alpha \equiv r^r \ln r$, in terms of which $\beta_0 = r^{1-r}$, as can be directly verified by substituting into the equation.  This implicit approximate solution is correct to order $1/\alpha^2$ for large $\alpha$.  If we try to produce an explicit solution from this, we run into  correction terms like $\log(\log(\alpha))$ and the convergence is super-slow. Nevertheless, a simple rough approximation
is
\begin{equation}
\beta_0 \approx \frac{\ln \alpha}{\alpha},
\end{equation}
up to corrections of $\ln(\ln r)/\alpha$, which is correct to better than $6\%$ for $\alpha>10$.
We can now approximate $F(\beta_0)$ for large $\alpha$, where the $\alpha\beta_0^2/2$ term is dominant, and so
\begin{equation}
F(\beta_0)\approx \frac{r^{2-r}\ln r}{2} \approx \frac{\ln^2 \alpha}{2\alpha}
\end{equation}

Our result connects directly with our previous result, Eq. (\ref{eqa0}), when $\alpha$ is ${\cal O} (1)$.  Remember the relation between $p$ and $\alpha$, as $\alpha$ becomes large, $p$ moves away from the region close to unity, and so
\begin{equation}
\ln SU(N) \approx \frac{\ln^2 N(1-p)}{2(1-p)} \approx \frac{\ln^2 N}{2 \ln (1/p)}
\end{equation}
as expected.

 In the opposite limit, when $\alpha$ is very small, say, $\alpha = \gamma/N$ (so $p = 1-\gamma/N^2$), $\beta_0 = 1-\alpha$, and the  exponential term of (\ref{eqa8}), $\exp(\gamma/2)$, is unity at $\gamma=0$ and grows exponentially with $\gamma$, as discussed above.

\begin{figure}
\includegraphics[width=7cm]{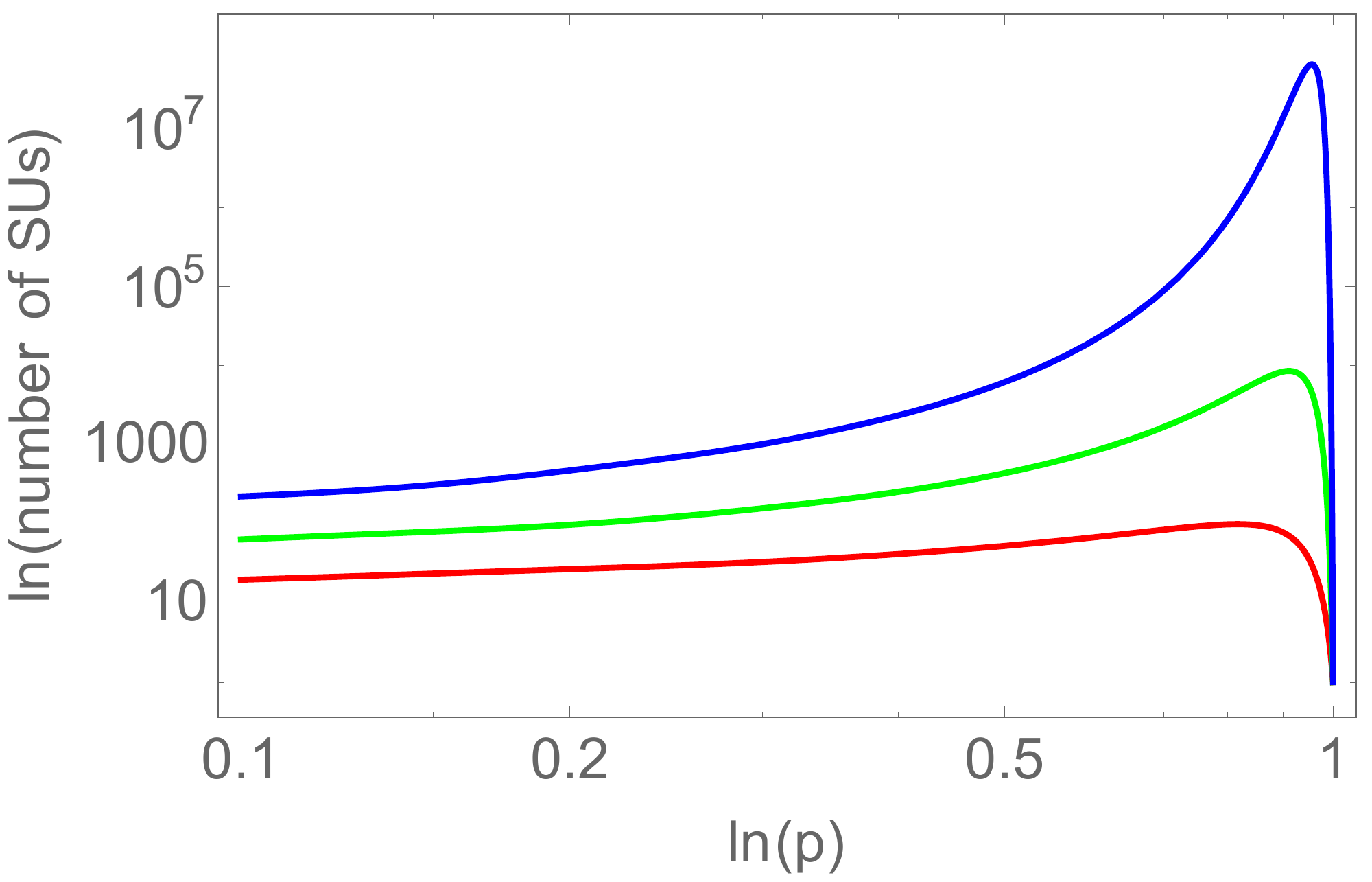}
\caption{$\ln(SU)$ vs. $\ln(p)$, as obtained from a numerical summation of the Bollob{\'a}s-Erd{\"o}s formula for the symmetric case, for $N=20$ (red) $N=40$ (green) and $N=80$ (blue). The growth of the number of SUs with $N$ becomes exponential when $1-p$  is ${\cal O} (1/N)$, as expected.  When $1-p$  is ${\cal O} (1/N^2)$, there is a drop towards one SU with all the $N$ species. In this regime the number of $SU$s is independent of $N$. }
\label{fig3}
\end{figure}

\begin{figure}
\includegraphics[width=9cm]{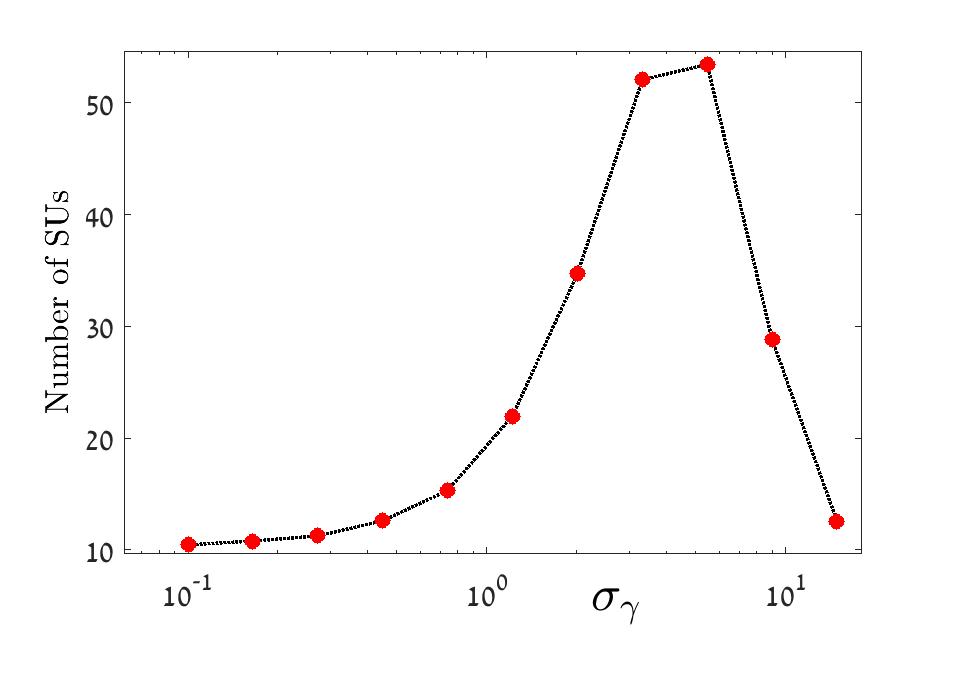}
\caption{Number of SUs (averaged over 500 samples) vs. $\sigma_\gamma$ as obtained from an exact enumeration of all the stable and uninvadable combinations of species in the Gamma model for $N=20$. Red points are actual results, each obtained from examination for stability and uninvadability  of all the $2^{20}$ combinations of species,  the dashed line is just to guide the eye.  The Gamma model is described by Eq. (\ref{eqa3a}) with $C=1$ and where each pair of numbers $c_{i,j}=c_{j,i}$ is picked independently from a Gamma distribution with mean one and standard deviation $\sigma_\gamma$. While the number of SUs is smaller than their number in the corresponding Binomial model (we have suggested in \cite{fried2016communities} that the Binomial model yields an upper bound for the Gamma) we still observe the growth in the number of SUs when $\sigma_\gamma \sim \sqrt{N}$ (around 4-5), then it drops towards the full coexistence phase.  }   \label{fig4}
\end{figure}

 These three regimes are depicted in Figure \ref{fig3}. As opposed to the case where $p$ is ${\cal O} (1)$ (or $\alpha$  is ${\cal O} (N)$) considered in \cite{fried2016communities}, where the growth is $N^{\ln N}$ type, when $\alpha$  is ${\cal O} (1)$ the growth is exponential while if $\alpha$ is ${\cal O} (1/N)$ the number of SUs is close to one. In figure \ref{fig4} we show that the same qualitative behavior is observed in the corresponding symmetric Gamma model~\cite{fried2016communities}.

\section*{The asymmetric Network}

Unlike the symmetric case where $c_{i,j}=c_{j,i}$, in an asymmetric system $c_{i,j}$ and $c_{j,i}$ are drawn independently from a given distribution. In this section we consider the Binomial model in this case.

The strong competition phase of the asymmetric Binomial interaction model allows for \emph{three} types of relationships between species $i$ and $j$. As in the symmetric case, it may happen that $c_{i,j} = c_{j,i} = 0$, so the species are non-interfering, and $c_{i,j} = c_{j,i} = A$, meaning that for large enough $A$ the two species are mutually exclusive. The asymmetry allows for a third, \emph{dominance}, relationship at large $A$: if $c_{i,j} = A, \  c_{j,i} = 0$, species $i$ may invade $j$ but the opposite process is forbidden. Accordingly,  $j$ may be a member of a maximal  clique only if another species in this clique is uninvadable by $i$.

In the asymmetric Binomial model, we define $\tilde{p}$ to be the chance that a \textbf{single entry} of the interaction matrix is zero. The argument presented at the beginning of the last section here yields  a completely different answer. Starting from $\tilde{p}=1$ and breaking one link (say, between 1 and 2), implies that 1 can invade 2 but 2 cannot invade 1, so the number of maximum cliques remains one. This will be the case until we hit both links between two specific species, and again this will happen only when $1-\tilde{p}$ is ${\cal O}(N)$. Accordingly, at the asymmetric case we expect that the number of SUs for $p$ close to one will be ${\cal O}(1)$.

By extending this argument, one can develop some intuition for the generic case which is neither symmetric nor asymmetric. In general one may expect that the stress species 1 suffers from 2 is not exactly the same as the stress species 2 suffers from 1, but that they are related to each other. For example, if there is some niche overlap between species 1 and 2, but the niche  of 1 is wider than the niche of 2, one expects $c_{1,2} < c_{2,1}$, but their values are correlated.

Naively, one may guess that symmetry is a ``fragile" property, so any deviation from perfect symmetry will send the system to the equivalence class of the asymmetric model. However, our argument allows us to realize that the opposite is true. As long as the system allows for a finite fraction of symmetric links, the breaking of each of them doubles the number of maximal cliques when the system is very close to its complete graph limit. Accordingly, as long as there is \emph{some} symmetry in the problem ($c_{i,j}$ is positively correlated with $c_{j,i}$) one should expect an exponentially large number of SUs when $\alpha$ is ${\cal O} (1)$, although the coefficient of $N$ in the exponent falls along with the degree of correlation. As we shall see below, the result in the purely asymmetric case reflects a ``miraculous" cancelation of terms, so this  turns out to be the fragile case.

 In Ref. \cite{fried2016communities}, we extended the  Bollob{\'a}s-Erd{\"o}s formula to the asymmetric case, showing that the number of SUs satisfies,
\begin{equation} \label{eqa20}
SU(N,S) = {N \choose S} \tilde{p}^{S(S-1)} (1-\tilde{p}^S)^{N-S}.
\end{equation}
Interestingly, the only difference between (\ref{eqa20}) and   (\ref{eqa1}) is the factor of 2 in the second term, reflecting the fact that for a collection of $S$ species to be noninterfering one needs all the $S(S-1)/2$  $c_{i,j}$s to be zero in the symmetric case, while in the asymmetric case $c_{i,j}$ and $c_{j,i}$ are picked at random so the number of independent links is doubled. As we shall see, this innocent looking modification has highly nontrivial consequences.

Implementing the method used for the symmetric case, one finds,
\begin{equation} \label{eqa21}
SU(N,S)=\frac{e^{NF_\textit{as}(\beta)}}{\sqrt{2\pi N}},
\end{equation}
where,

\begin{eqnarray} \label{eqa22}
F_\textit{as}(\beta)&=&-(1-\beta)\ln(1-\beta)-\beta \ln(\beta) \\ \nonumber &-&\alpha\beta^2+(1-\beta)\ln(1-e^{-\alpha\beta}).
\end{eqnarray}

As before, this pair of formulas appear to suggest  that, as long as both $\alpha$ and $\beta$ are ${\cal O}(1)$, the number of SUs is exponential in $N$. However, we shall see that in this case $F_\textit{as}(\beta_0)=0$ and the actual large-N asymptotic turns out to be non-exponential.

The equation for $\beta_0$ now reads

\begin{eqnarray} \label{eqa23}
F'_\textit{as}(\beta_0)&=& -2\alpha\beta_0-\frac{\alpha(-1+\beta_0)}{-1+e^{\alpha\beta_0}} +
\ln(1-\beta_0) \\ \nonumber &-&\ln(\beta_0)-\ln(1-e^{-\alpha\beta_0}) = 0.
\end{eqnarray}

Surprisingly, one can find an \emph{exact} solution to this equation,
\begin{equation} \label{eqa24}
\beta_0 = \frac{W(\alpha )}{\alpha },
\end{equation}
where $W$ is the Lambert W function, defined by $W(x)\exp[W(x)]=x$.
 Plugging this into $F$, we find that $F_\textit{as}(\beta_0)$ \emph{vanishes identically}. This implies that the first contribution from the Laplace integral is ${\cal O}(1)$ (instead of being exponential in $N$) so we should repeat the exercise from its starting point, keeping all the ${\cal O}(1)$ terms (omitting only ${\cal O}(1/N)$ and other small terms).

$F_\textit{as}$ in the controlling factor of (\ref{eqa21}) then takes the form,

\begin{eqnarray}
F_\textit{as}&=&-(1-\beta+\frac{1}{2N})\ln(1-\beta)-(\beta+\frac{1}{2N})\ln(\beta)
\\ \nonumber &-&\alpha\beta^2+\frac{\alpha\beta}{N}
-\frac{\alpha^2\beta^2}{2N}+(1-\beta)\ln(1-e^{-\alpha\beta-\frac{\beta\alpha^2}{2N}}).
\end{eqnarray}

To continue, let us write $F_\textit{as}$ as,
\begin{equation}
F_\textit{as}=F_\textit{as}^{(0)} +\frac{F_\textit{as}^{(1)}}{N}
\end{equation}
where $F_\textit{as}^0$ is given in (\ref{eqa22}) and
\begin{widetext}
\begin{equation}
F_\textit{as}^{(1)} = \frac{\alpha ^2 \beta ^2}{2}+\frac{\alpha ^2 \beta ^2}{2 \left(e^{\alpha  \beta }-1\right)}-\frac{\alpha ^2 \beta }{2 \left(e^{\alpha  \beta }-1\right)}-\alpha  \beta +\frac{1}{2} \log (1-\beta )+\frac{\log (\beta )}{2}.
\end{equation}
\end{widetext}
Evidently, the main contribution in the large $N$ limit still comes from $\beta_0$ given in (\ref{eqa24}) and,
\begin{equation}
e^{F^{(1)}_\textit{as}(\beta_0)} = \frac{1}{2} \left(\log \left(1-\frac{W(\alpha )}{\alpha }\right)-3 W(\alpha )\right),
\end{equation}
meaning that there is no exponential growth of the number of maximal cliques with $N$.

Now we can implement the Laplace integral scheme to (\ref{eqa21}) (the sum over $S$ is converted to an integral over $N d\beta$) to obtain,
\begin{widetext}
\begin{equation} \label{eqa30}
SU(N)=N\int_0^1 d \beta \frac{e^{NF_\textit{as}(\beta)}}{\sqrt{2\pi N}} = \frac{N}{\sqrt{2 \pi}} e^{F^{(1)}_\textit{as}(\beta_0)} \int_{-\infty}^{\infty} e^{-N|(F^{(0)}_\textit{as})''|_{\beta_0}|(\beta-\beta_0)^2}= \frac{e^{F^{(1)}_\textit{as}(\beta_0)}}{\sqrt{2|(F^0_\textit{as})''|}},
\end{equation}
\end{widetext}
so
\begin{equation} \label{eqa31}
SU(N)= \frac{\alpha}{W(\alpha)+W^2(\alpha)}
\end{equation}

\begin{figure}
\includegraphics[width=7cm]{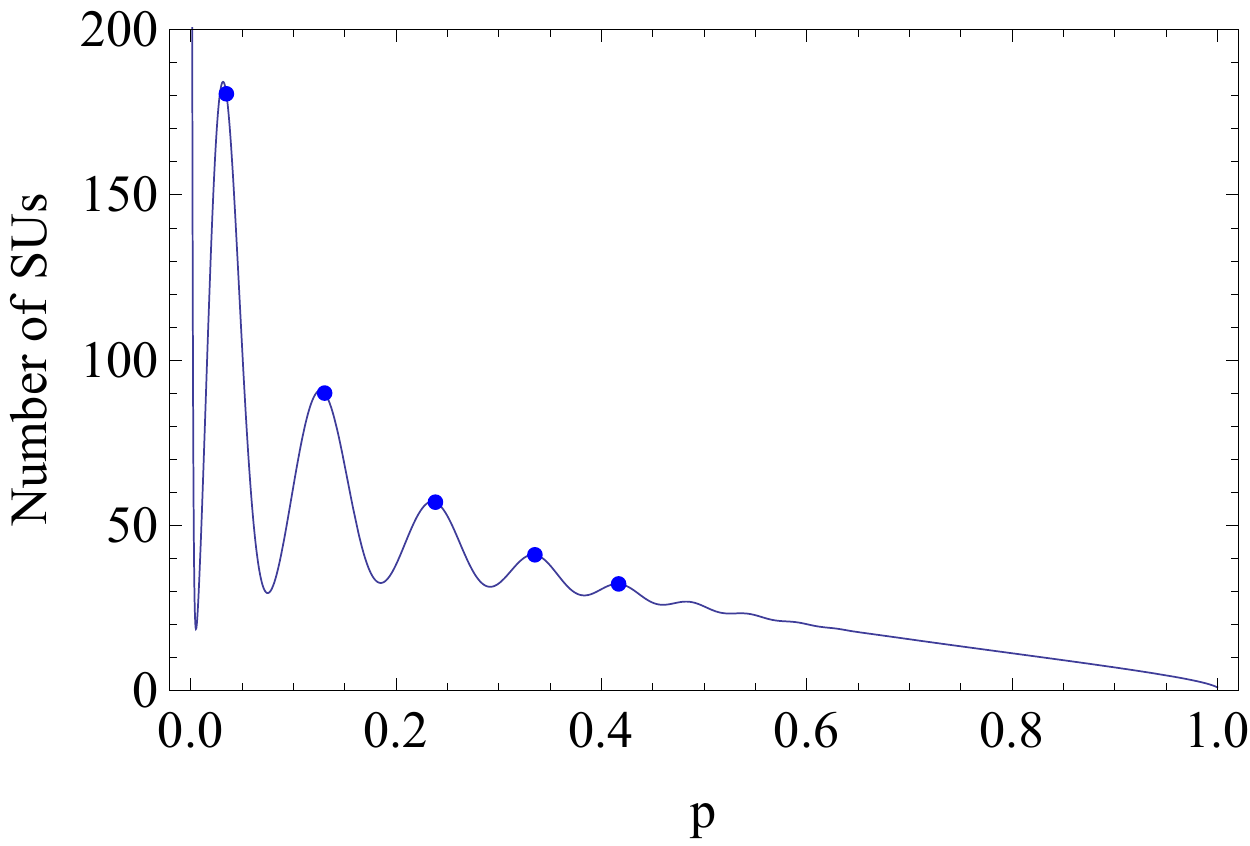}
\caption{$\ln(SU)$ vs. $p$, as obtained from a numerical summation of the Bollob{\'a}s-Erd{\"o}s formula for the asymmetric case, for $N=1000$. In general the number of SUs decays with $p$, with no exponential peak close to the fully connected limit. This general trend is superimposed on oscillations in the region where $p$ is ${\cal O} (1)$, since in this regime there is only one integer value of maximal clique sizes that dominate the sum, as explained in the text. To demonstrate that, thick  points were added to mark the $p$ values for which $S = \beta_0 N$ is $2,3,4,5$ and $6$.  }\label{fig5}
\end{figure}

In the limit where $\alpha$ is ${\cal O} (N)$  considered  in \cite{fried2016communities} the width of the Gaussian in the integration of (\ref{eqa30}) is $1/N$, meaning that only a single large term in  the sum of $SU(N,S)$ over $S$ contributes (see Fig. \ref{fig5}). In such a case there is no contribution from the integration around the maximum, and the number of cliques is
\begin{equation}
SU(N) \sim \frac{e^{F^{(1)}_\textit{as}(\beta_0)}}{\sqrt{2 \pi N}} \sim \frac{N}{\ln^{3/2}(N)}
\end{equation}
as shown in \cite{fried2016communities}.

\begin{figure}
\includegraphics[width=9cm]{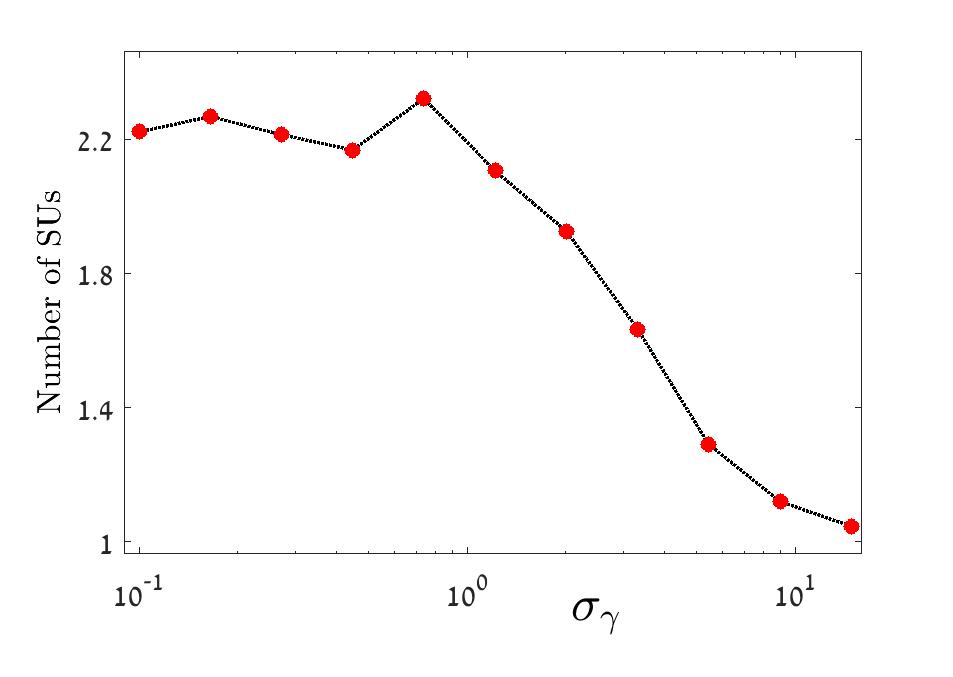}
\caption{Number of SUs (averaged over 500 samples) vs. $\sigma_\gamma$ as obtained from an exact enumeration of all the stable and uninvadable combinations of species in the asymmetric Gamma model for $N=20$. Red points are actual results, each obtained from examination for stability and uninvadability  of all the $2^{20}$ combinations of species,  the dashed black line is just to guide the eye.  The asymmetric Gamma model is described by Eq. (\ref{eqa3a}) with $C=1$ and where each $c_{i,j}$ is picked independently from a Gamma distribution with mean one and standard deviation $\sigma_\gamma$.  }   \label{fig6}
\end{figure}

The behavior of the number of SUs in different regimes of the Binomial model is depicted in Figure \ref{fig5}, and the results of the corresponding Gamma model are shown in Figure \ref{fig6}.

\section{Discussion}

In this paper we have studied  the number of stable and uninvadable states in an ecological network. We assumed a local community which is coupled to a regional species pool. In the local community the particular level of competition between any given pair of species was drawn at random.

Many empirical ecological networks (in particular food webs \cite{drossel2001influence} and networks with mutualistic interactions \cite{suweis2013emergence}) were shown to admit a nontrivial structure (like modularity or nestedness) so their description as random networks is problematic. Still, we believe that the analysis presented here is relevant to various aspects of the general problem. First, there are less evidence, as far as we know, for a general structure in systems of competing species (see, e.g., \cite{volkov2009inferring}). Second, even if the mainland interactions are structured, there is no a priori reason to assume that this is the case on the island. A third point (which is complementary to the second) is that, when the interactions are inferred from empirical studies of local communities, one would like to understand what aspects of these interactions are the result of the restriction of a regional system with a (possibly) different structure to its SUs.

The model considered here is characterized by three parameters: the number of species in the regional pool $N$, the mean value of the off-diagonal entries of the competition matrix $C$  and the parameter that reflects the heterogeneity of the competition terms, $\tilde{\sigma}^2$.  In the Binomial model $\tilde{\sigma}^2 = C^2 p/(1-p)$, so in the limit when $p = 1-\alpha/N$, $$\tilde{\sigma}^2 \sim \frac{C^2 N}{\alpha}.$$

 In the works of Mehta and Fisher \cite{fisher2014transition} and Bunin \cite{bunin2016interaction} the average value of an (off diagonal) interaction matrix term and the variance of these terms both are taken to be of order $1/N$. This parameter regime is right on the border of the regime defined by May's stability criteria mentioned above.  Translating this to the notations of our paper, one has $C \sim \mu/N$, say (when $\mu$ is ${\cal O} (1)$), hence $\sigma^2 = \mu^2/N\alpha$. So the regime of parameters covered by the $p = 1-\alpha/N$ limit of the Binomial model (with $\alpha$ order one) includes the regime considered in  \cite{fisher2014transition,bunin2016interaction} as a special case.

 The main outcome of the analysis presented here and in \cite{fried2016communities} is that the number of SUs grows exponentially with $N$ \emph{if} $\alpha$ is ${\cal O} (1)$, and the matrix in not purely asymmetric. If $p$ is order one, or in the case of an asymmetric competition terms, the growth is subexponential, ranging from $N^{\ln(N)}$ dependency to sublinear growth. This implies that, as long as an exponential number of SUs is required for a glass transition (as suggested by the analogy with the random energy model presented in \cite{fisher2014transition}), such a transition occurs only in the regime of very weak competition and very large systems.

 When ecologists  consider high-diversity assemblages and  try to understand the forces that shape their structure, they usually have in mind  systems  like tropical trees~\cite{ter2013hyperdominance}, coral reef~\cite{connolly2014commonness} or plankton~\cite{stomp2011large}. In these cases the level of niche overlap between species is evidently quite high, as all these species are using the same set of a few key resources, more or less in the same manner. Accordingly, one should expect these systems to be in the regime where the interaction terms of the competition matrix are ${\cal O} (1)$ (see, e.g. the recent study \cite{carrara2015experimental}), where the number of species in an SU scales logarithmically with $N$ \cite{fried2016communities}, the number of SUs is subexponential, and there is no glass transition.

 To the best of our understanding, the parameter regime considered in \cite{fisher2014transition,bunin2016interaction} and here corresponds to a completely different scenario. This is the case of a community with many species but with strong niche partitioning (say, many bird species with different beak size, eating different kinds of food) that still have weak competition between species (due to some overlap in the type of food they are eating, weak  nest site competition or due to predation by a common predator). Most ecologist feel that the coexistence of many different species in such a scenario needs no explanation (since the main issue they consider is the competitive exclusion principle) but in fact there is still a theoretical problem, namely May's complexity-diversity relationship, meaning that even a community with very weak interactions will collapse when the number of species is large. Here we have shown that in this case one should expect to see a local community with ${\cal O} (N)$ species ($\beta_0$ is order one), and possibly some kind of a glass transition. The relevance of this theoretical framework to empirical systems appears to be an open problem.

\bibliography{references1}

\end{document}